\documentclass[a4paper]{VisionStyle}
\usepackage{epsfig}

\newcommand{\gtae}{$\buildrel {\lower3pt\hbox{$>$}} \over 
{\lower2pt\hbox{$\sim$}} $}
\newcommand{\ltae}{$\buildrel {\lower3pt\hbox{$<$}} \over
{\lower2pt\hbox{$\sim$}} $}

\begin{document}

\title{State transitions in LMC X-3}

\author{R.\,Soria\inst{1}, M.\,J.\,Page\inst{1} \and K.\,Wu\inst{1}} 

\institute{Mullard Space Science Laboratory, 
          University College London, Holmbury St Mary, 
          Surrey RH5 6NT, UK }

\maketitle 

\begin{abstract}

We carried out a multiwavelenght 
study of the black-hole candidate LMC X-3 with {\it XMM-Newton}. 
The system showed a transition to a low-hard state, in which 
the X-ray spectrum was well fitted by a simple power law. It then 
returned to a high-soft state, characterised by a strong 
disk-blackbody component. The line-of-sight absorption 
column density is \ltae~$4 \times 10^{20}$ cm$^{-2}$, 
consistent with the foreground Galactic absorption. 
This rules out wind accretion. We argue that, despite LMC X-3 being 
a high-mass X-ray binary, Roche-lobe overflow 
is the main mechanism of mass transfer. 
From UV/optical observations in the low-hard state, we determine 
that the companion is a slightly evolved B5 star with 
a mass $M \approx 4.5$ M$_{\sun}$. 
This is indeed consistent with the secondary 
star being close to filling its Roche lobe.

\end{abstract}

\section{Introduction}

LMC X-3 (\cite{rsoria-C1:le71}) is a persistent X-ray source in the 
  Large Magellanic Cloud (LMC).
The orbital period is 1.705~d 
(\cite{rsoria-C1:va85}; \cite{rsoria-C1:va87}). 
The non-detection of eclipses in the X-ray curve    
  implies that the orbital inclination of the system 
  is \ltae~$70^{\circ}$ (\cite{rsoria-C1:co83}).  
Its optical brightness ($V \sim 17$) 
  indicates that the system has a massive companion.
From optical spectroscopic observations, 
its mass function was estimated to be $\simeq 2.3$ M$_{\odot}$ 
(\cite{rsoria-C1:co83}). The inferred mass 
of the compact object in LMC X-3 would then be \gtae~$7$ M$_{\odot}$ 
  (\cite{rsoria-C1:pa83}). 
If the effect of soft X-ray irradiation 
on the surface of the secondary star is taken into account, 
a mass function $f_{\rm M} = 1.5 \pm 0.3$ M$_{\odot}$ is obtained instead 
(\cite{rsoria-C1:so01}). This corresponds 
to a lower limit for the mass of the compact object 
$M_{\rm X} > (5.8 \pm 0.3)$ M$_{\odot}$.
Thus, the system is a black-hole candidate (BHC).

\section{Spectral states and mass transfer mechanism}

Most BHC show transitions between soft 
and hard X-ray spectral states. In the soft state, their X-ray spectrum 
consists of a thermal component and a power-law component; 
in the hard state, the thermal component 
is insignificant and the power law is harder. 
The thermal component, which can be fitted by a blackbody 
or disk-blackbody spectrum with a temperature $\sim 1$~keV, 
is interpreted as thermal emission from the inner accretion disk.
The power-law component is believed to be Comptonised emission 
  from a disk corona (\cite{rsoria-C1:su00}) 
  or from the high-speed infalling plasma  
  near the black-hole event horizon (\cite{rsoria-C1:ti98}). 
The photon index of the power law  $\Gamma \approx 2.5$--4 
in the soft state, and $1.5$ \ltae~$\Gamma$ \ltae~$2$ in the hard state.

\begin{figure}[t]
  \begin{center}
    \epsfig{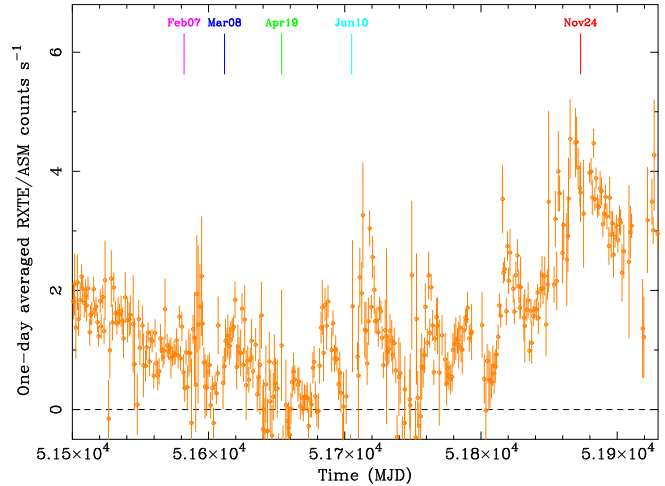}
  \end{center}
\caption{The time of our {\it XMM-Newton} observations of LMC X-3, 
            marked by solid vertical lines,            
            are plotted above 
            the {\it RXTE}/ASM light curve. }  
\label{rsoria-C1_fig:fig1}
\end{figure}

LMC X-3 is normally found in the soft state. 
A rare transition to the hard state occurred in 2000 April--May  
(\cite{rsoria-C1:bo00}). Other short-duration  
transitions to a hard state were observed 
in 1997 and 1998 (\cite{rsoria-C1:wi01}).
LMC X-1, another high-mass BHC in the LMC, 
  has been seen in the soft state only.   
In contrast, Cyg X-1, the high-mass BHC in our Galaxy, 
  tends to be in the hard state for the majority of the time. 

An unsolved problem for LMC X$-$3 
is the process of mass transfer. With an estimated
mass of the companion star $4$ M$_{\odot}$ \ltae~$M_2$ \ltae~$8$ M$_{\odot}$ 
(\cite{rsoria-C1:co83}), 
the system appears to be intermediate between high-mass black-hole binaries 
such as Cyg X$-$1 (mass of the companion star $\approx 33$ M$_{\odot}$, 
see \cite{rsoria-C1:gi86}), 
and low-mass black-hole binaries such as A0620$-$00 
(mass of the companion star $\approx 0.7$ M$_{\odot}$). 
In the former class of systems, 
mass transfer occurs mainly via a stellar wind, and the donor star 
is more massive than the primary; in the latter, the donor star 
is usually a late-type star filling its Roche lobe.
We used {\it XMM-Newton} to study the spectral behaviour 
of LMC X-3 over its spectral state transition, 
and to determine the mechanism of mass trasfer in this system.

\section{Results of our {\it XMM-Newton} study}    
\label{rsoria-C1_sec:stu}
 
LMC X-3 was observed with 
  the European Photon Imaging Camera (EPIC), 
  the Reflection Grating Spectrograph (RGS) and  
  the Optical Monitor (OM) on board {\it XMM-Newton}, 
  between 2000 February and November. See also 
\cite*{rsoria-C1:wu01} and \cite*{rsoria-C1:so01}. 
All EPIC exposures were taken with the ``medium'' filter; 
Some are affected by pile-up due to the high count rate. 
Here, we present only the EPIC-PN exposures not affected 
by pile-up (ie., those taken in ``small window'' 
and in ``timing'' mode; see the {\it XMM-Newton} Remote Proposal 
    Submission Software Users' Manual).  
The log of the EPIC-PN observations used for this study 
  is shown in Tables 1; the log of the RGS observations 
is listed in Table 2.
The data were processed using the 5.1 version of the SAS.

\begin{table}[t]
%\begin{table*}
%\begin{minipage}{150mm}
%\begin{minipage}{80mm}
\caption{{\it XMM-Newton} EPIC-PN Observation Log}
\label{mathmode} 
  \begin{center}
    \leavevmode
    \footnotesize
%\centering 
\begin{tabular}{@{}lccr} 
\hline\\[-5pt]
  Rev.
%&  Instrument
%\footnote{Here and in the following tables, 
%   ``s'' and ``u'' label  
%    ``scheduled'' and ``unscheduled'' 
%   PV exposures during the same revolution}
  & Start--end (MJD)  
  & Live exp. time
%   \footnote{For the PN small window mode, the live time was 71\% 
%of the exposure time listed here; for the MOS RFS mode the live time 
%was 6.9\% of the exposure time.}   
  & Mode 
%   \footnote{See {\it XMM} Remote Proposal 
%    Submission Software Users' Manual}  
  \\ [5pt]
\hline \\[-5pt]
%  Rev0028/201 &  PN    
%& 2000/02/02 14:01:41 -- 2000/02/02 20:04:14   
%                       & 08.1~ks    & small window \\[2pt]             
%  Rev0028/301 &  PN  
%& 2000/02/02 20:04:14 -- 2000/02/02 20:21:32   
%                       & 0.3~ks    & full PN      \\[2pt] 
%          &  MOS1   
%& 2000/02/03 00:21:22 -- 2000/02/03 03:06:21   
%                       & 08.9~ks    & partial RFS  \\[2pt] 
%  Rev0030/501 & MOS1  
%& 2000/02/07 23:52:21 -- 2000/02/08 01:41:34   
%                       & 05.9~ks    & partial W5   \\[2pt] 
%  Rev0041/101 &  PN     
%& 2000/02/28 14:41:53 -- 2000/02/28 15:14:42   
%                       & 0.7~ks    & full PN       \\[2pt]
%  Rev0041/401 & MOS1  
%& 2000/02/28 18:58:49 -- 2000/02/28 20:13:18   
%                       & 03.5~ks    & partial W2   \\[2pt] 
%          &  MOS1   
%& 2000/02/28 23:07:23 -- 2000/02/29 00:16:48 
%                       & 03.6~ks    & partial W5   \\[2pt]             
%          &  MOS1   
%& 2000/02/29 01:11:39 -- 2000/02/29 10:01:16   
%                       & 03.4~ks    & partial W4   \\[2pt]   
%  Rev0045/101 &  PN    
%& 2000/03/08 13:49:59 -- 2000/03/08 14:18:03   
%                       & 01.6~ks    & full PN      \\[2pt]             
%  Rev0045/201 &  PN      
%& 2000/03/08 16:36:22 -- 2000/03/08 19:39:15   
%                       & 11.0~ks    & ext.\ fullframe      \\[2pt]    
%  Rev0045/301 & MOS1  
%& 2000/03/08 18:29:08 -- 2000/03/08 20:20:54   
%                       & 06.6~ks    & partial W3   \\[2pt]    
  0066 
%& 2000/04/19 12:57:04 -- 2000/04/19 20:20:22   
& 51653.5396--51653.8475
% 26.6~ks 
                       & 05.8~ks 
%                       \footnote{Only the first 8.3~ks are useful}    
                                    & small w \\[5pt]    
  0092  
%& 2000/06/10 00:13:47 -- 2000/06/10 06:57:07   
& 51705.0096--51705.2897
% 24.2~ks
                       & 16.9~ks    & small w \\[5pt]        
  0176  
%& 2000/11/24 23:02:21 -- 2000/11/25 04:17:09   
& 51872.9600--51873.1786
                       & 09.2~ks    & timing \\[5pt]
%            &  MOS1  
%& 2000/06/10 00:53:03 -- 2000/06/10 06:54:43  
%                       & 20.3~ks    & partial W4 \\[2pt] 
\hline    
%\end{tabular}   
%\end{minipage} 
%\end{table*} 
      \end{tabular}
  \end{center}
\end{table}  

\begin{table}[t]
%\begin{table*}
%\begin{minipage}{150mm}
%\begin{minipage}{80mm}
\caption{{\it XMM-Newton} RGS Observation Log }
\label{mathmode2} 
  \begin{center}
    \leavevmode
    \footnotesize
%\centering 
\begin{tabular}{@{}lccr} 
\hline\\[-5pt]
   Rev. 
  &  Start--end (MJD)  
  & Exp. time 
%  & count rate \footnote{In photons~s$^{-1}$} 
& Instrument
  \\[5pt] 
\hline \\[-5pt]
%  0028   
%&  2000/02/02 13:42:17 -- 2000/02/02 18:36:36   
%& 51576.5710--51576.7754
%                       & 17.6~ks 
%& RGS2
%& 4.79$\pm$0.02 \\[2pt] 
  0030 
%& 2000/02/07 22:50:33 -- 2000/02/08 02:21:54   
& 51581.9517--51582.0985
                       & 12.1~ks 
%& 4.52$\pm$0.03 
& RGS2\\[5pt]
  0045
& 51611.7275--51611.8367
& 09.0~ks
& RGS1+2\\[5pt]
%   0045              
%&  2000/03/08 13:00:10 -- 2000/03/08 14:48:41   
%& 51611.5418--51611.6171
%                       & 06.4~ks & 3.59$\pm$0.03 \\[2pt]  
%& RGS2 (s1) 
%&  2000/03/08 13:00:10 -- 2000/03/08 14:48:38   
%                       & 06.4~ks & 3.56$\pm$0.03 \\[2pt]            
%  Rev0045/201 & RGS1 (s1) 
%&  2000/03/08 15:13:58 -- 2000/03/08 17:07:28   
%& 51611.6346--51611.7135
%                       & 06.7~ks & 3.62$\pm$0.02 \\[2pt] 
%            & RGS2 (s1) 
%&  2000/03/08 15:13:58 -- 2000/03/08 17:07:28
%& 51611.6346--51611.7135
%                       & 06.8~ks & 3.57$\pm$0.02 \\[2pt] 
%  Rev0045/301 & RGS1 (s1) 
%&  2000/03/08 17:27:39 -- 2000/03/08 18:19:32   
%& 51611.7275--51611.8367
%                       & 03.0~ks & 3.66$\pm$0.04 \\[2pt] 
%            & RGS1 (s3) 
%&  2000/03/08 18:20:16 -- 2000/03/08 19:12:12 
%                       & 03.0~ks & 3.57$\pm$0.04 \\[2pt]   
%            & RGS1 (s5) 
%&  2000/03/08 19:12:53 -- 2000/03/08 20:04:49 
%                       & 03.0~ks & 3.40$\pm$0.04 \\[2pt]     
%            & RGS2 (s2) 
%&  2000/03/08 18:20:16 -- 2000/03/08 18:26:18  
%                       & 00.3~ks & 3.63$\pm$0.17 \\[2pt]  
%            & RGS2 (s4) 
%&  2000/03/08 17:27:39 -- 2000/03/08 18:19:32  
%                       & 03.0~ks & 3.26$\pm$0.04 \\[2pt]   
%            & RGS2 (s6) 
%&  2000/03/08 19:12:53 -- 2000/03/08 20:04:49  
%                       & 00.3~ks & 3.18$\pm$0.17 \\[2pt]   
  0066             
& 51653.1377--51653.6587
& 44.5~ks
& RGS1+2 \\[5pt]
%&  2000/04/19 03:18:19 -- 2000/04/19 15:48:34  
%& 51653.1377--51653.6587
%                       & 44.3~ks & 0.0154$\pm$0.0012 \\[2pt]  
%& RGS2 (s8) 
%&  2000/04/19 03:18:19 -- 2000/04/19 15:48:33   
%                       & 44.3~ks & 0.0153$\pm$0.0012 \\[2pt]  
  0092             
& 51704.2877--51705.2917
& 77.6~ks
& RGS1+2 \\[5pt]
%&  2000/06/09 06:54:19 -- 2000/06/09 21:22:49  
%& 51704.2877--51704.8908
%                       & 51.9~ks & 1.32$\pm$0.01 \\[2pt]  
%& RGS2 (s2) 
%&  2000/06/09 06:54:19 -- 2000/06/09 21:22:48    
%                       & 51.9~ks & 1.53$\pm$0.01 \\[2pt]            
%  Rev0092/201 & RGS1 (s4) 
%&  2000/06/09 23:51:39 -- 2000/06/10 07:00:04   
%& 51704.9942--51705.2917
%                       & 25.6~ks & 1.56$\pm$0.01 \\[2pt] 
%            & RGS2 (s5) 
%&  2000/06/09 23:51:39 -- 2000/06/10 07:00:09   
%                       & 25.6~ks & 1.34$\pm$0.01 \\[2pt]           
  0176
& 51872.9347--51873.1786
& 21.0~ks
& RGS1+2 \\[5pt]
\hline
\end{tabular}  
\end{center}
%\end{minipage} 
%\end{table*} 
\end{table}

\begin{figure}[!t]
  \begin{center}
    \epsfig{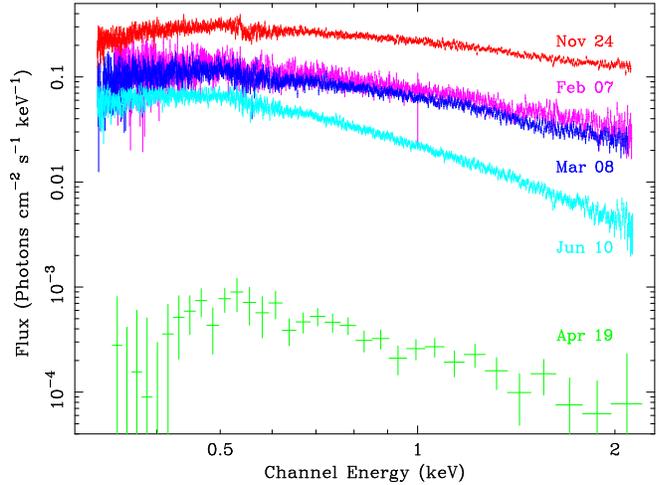}
  \end{center}
\caption{Unfolded {\it XMM-Newton}/RGS spectra show the transition 
from the high-soft to the low-hard state, and back to the high-soft 
state, between 2000 February and November.}  
\label{rsoria-C2_fig:fig2}
\end{figure}

\begin{table}[!t]
\caption{Spectral parameters of the {\it XMM-Newton} EPIC-PN observations}
\label{mathmode3} 
  \begin{center}
    \leavevmode
    \footnotesize
\begin{tabular}{@{}lccr} 
\hline\\[-5pt]
  Date of obs
& $T_{in}$ (keV)
& $\Gamma$
& $L_{0.3-10}$ (erg s$^{-1}$)\\[5pt]
\hline\\[-5pt]
Apr 19 
& not detected
& $1.9^{+0.1}_{-0.1}$
& $4.6 \times 10^{35}$\\[5pt]
Jun 10 
& $0.25^{+0.01}_{-0.01}$
& $1.89^{+0.02}_{-0.02}$
& $2.8 \times 10^{37}$\\[5pt]
Nov 24 
& $1.37^{+0.01}_{-0.01}$
& $2.59^{+0.05}_{-0.04}$
& $6.4 \times 10^{38}$\\[5pt]
\hline
\end{tabular}  
\end{center} 
\end{table}

The X-ray luminosity of LMC X-3 appeared to be declining    
  during our 2000 February--March observations, 
  with the {\it RXTE}/ASM $1.5-12$~keV count rate 
  generally below 2~ct~s$^{-1}$ (Figure~1).    
The Rev0066 observation (2000 April 19) was carried out 
  around the middle of a faint-hard state, 
  when the {\it RXTE}/ASM count rate was consistent with zero.   
The {\it RXTE}/PCA data obtained on May 5.76 and 10.01~UT 
 showed power-law spectra with a photon index $1.60\pm 0.05$  
  and a soft ($2$--$10$~keV) X-ray flux 
  of $\approx 5$--$9\times 10^{36}~{\rm erg}~{\rm s}^{-1}$ 
  at 50~kpc (\cite{rsoria-C1:bo00}).   
The system seemed to be returning to the high-soft state 
  at the time of the Rev0092 (June 10) observations. 
It was in the high-soft state during our last observation 
(Rev0176, November 24).

The thermal disk component disappeared in the low-hard state, 
but became dominant again as the system returned to the high-soft state 
(Table 3 and Figure 3). 
%The inner-disk temperature changed as shown in the figures below. 
The emitted luminosity in the 0.3-10 keV band varied by 3 orders 
of magnitude, reaching $L_{\rm x} \approx 6 \times 10^{38}$~erg~s$^{-1}$ 
(0.3--10.0~keV band) in November 2000 (here we have assumed  a total 
a column density $n_{\rm H} =  4 \times 10^{20}$~cm$^{-2}$ 
and a distance to the LMC 
of 50 kpc). This is the highest X-ray luminosity 
ever measured for this system.
The optical/UV luminosity increased by a factor 
of 2 (0.8 mag) in the high-soft state.

%$L_{\rm x} \approx 6 \times 10^{37}$~erg~s$^{-1}$ (0.3--8.0~keV band).   

\begin{figure}[!h]
  \begin{center}
    \epsfig{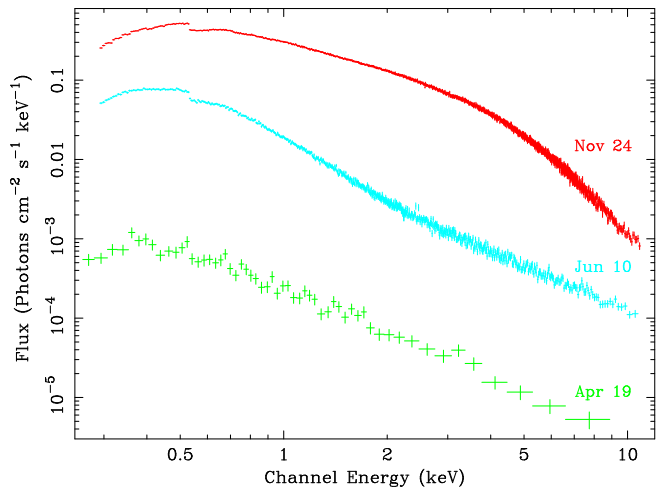}
  \end{center}
\caption{Unfolded {\it XMM-Newton}/EPIC-PN spectra show 
that the disk-blackbody component is not detected 
in the X-ray spectrum during the low-hard state (April 19). 
The spectrum in the low-hard state is a simple power-law. 
The disk-blackbody component becomes prominent again as LMC X-3 returns 
to the high-soft state. }  
\label{rsoria-C1_fig:fig3}
\end{figure}

\begin{figure}[ht]
  \begin{center}
    \epsfig{file=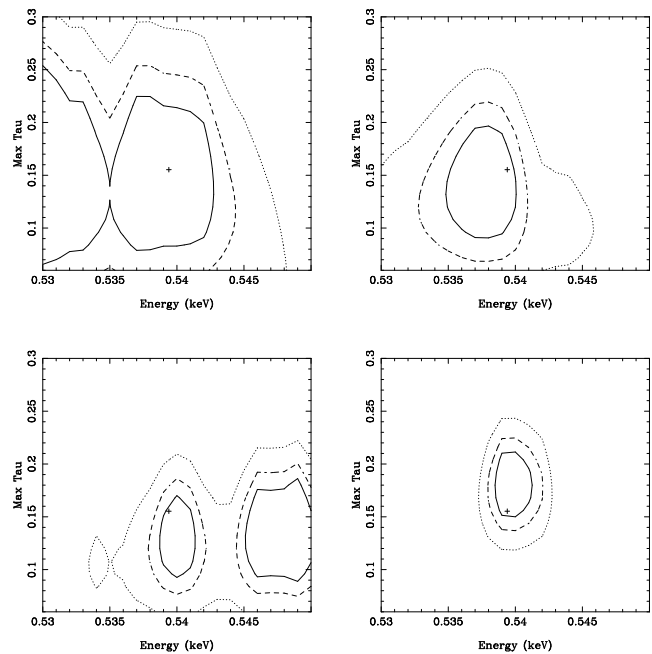, width=8.5cm}
  \end{center}
\caption{Results of our fitting a power law $\times$ edge 
model to a 3-\AA~region around the O\,I edge 
at $23$ \AA~($E = 0.54$ keV). 
The 1-$\sigma$, 2-$\sigma$ and 3-$\sigma$ confidence contours on the
wavelength and optical depth of the edge are shown for the spectra taken in
Rev. 30 (top left), 45 (top right), 92 (bottom left) and 176 (bottom right). 
The cross in each panel marks the best-fit parameters when all four
spectra are fitted simultaneously. Each of the four separate observations 
is consistent with this value. We infer that the total line-of-sight 
absorption is consistent with the expected Galactic 
interstellar absorption in the direction of LMC X-3.}  
\label{rsoria-C1_fig:fig4}
\end{figure}

The high-resolution RGS spectra allowed us to determine the absorbing 
column density for the X-ray emitting region. 
From the depth of the O\,I absorption edge at 
$23$ \AA~($E = 0.54$ keV), we find a total line-of-sight 
column density $n_{\rm H}$ \ltae~$4 \times 10^{20}$~cm$^{-2}$ (Figure 4). 
The foreground Galactic interstellar absorption 
in the direction of LMC X-3 is $n_{\rm H} = 3.2 \times  10^{20}$~cm$^{-2}$ 
(\cite{rsoria-C1:wi01}). Hence, the intrinsic column density 
is \ltae~$10^{20}$~cm$^{-2}$.  
This holds both in the low-hard and in the high-soft state, 
and is in contrast to the larger intrinsic column density    
  expected for a companion with a strong stellar wind. 

The non-detection of obvious emission lines 
  in the RGS spectra also indicates 
  the absence of wind matter ejected in previous epochs 
  (cf. the P Cygni lines seen in Cir X-1, \cite{rsoria-C1:br00}). 
Thus, the {\it XMM-Newton} spectral data rule out 
wind accretion from a massive companion. 
The high luminosity observed in the X-ray bands therefore 
  requires the companion to overflow its Roche lobe.

\section{Mass and spectral type of the companion star}
    
Observations of LMC X-3 in its X-ray low state allowed us to determine 
the mass and spectral type of the companion. The system was observed 
with {\it XMM-Newton}/OM on 2000 April 19. We obtained
an average brightness 
$v = 17.48 \pm 0.02$, $b = 17.39 \pm 0.02$, $u = 16.56 \pm 0.02$ 
in the three {\it{XMM-Newton/OM}} optical bands.
Using the latest available matrix of colour 
transformation coefficients (SAS Version 5.1, file OM\_COLORTRANS\_0005.CCF), 
we find that this corresponds to 
$V = 17.48 \pm 0.03$, $B = 17.36 \pm 0.03$, $U = 16.79 \pm 0.03$.
This implies a temperature $15500$ \ltae~$T_{\rm{eff}}$ 
\ltae~$16500$ and a spectral type B5 (Figure 5; 
see also \cite{rsoria-C1:so01}).
Hence, we infer that the companion is a slightly evolved star 
of mass $4.5$ \ltae~$M_2$ \ltae~$5.0$ M$_{\odot}$. 
No significant wind is expected from such a star, 
in agreement with the low column density 
inferred from the X-ray data.

\begin{figure}
  \begin{center}
    \epsfig{file=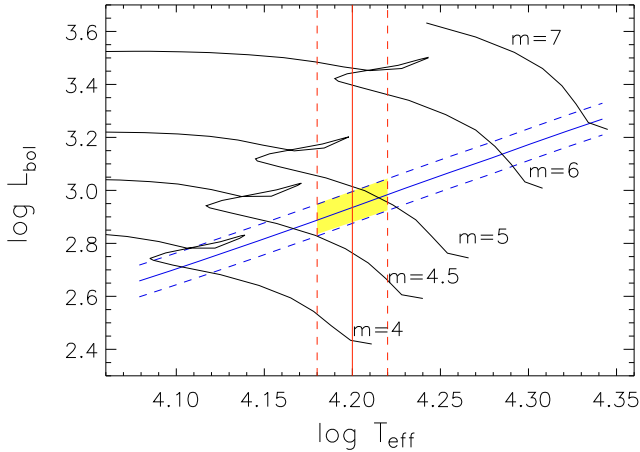, width=6.4cm, angle=270}
  \end{center}
\caption{The evolutionary tracks for stars of various masses, 
at Z = 0.008, in the 
($\log T_{\rm eff}$,\,$\log L_{\rm bol}$) plane, show the acceptable 
range of temperatures and luminosities derived from our {\it{XMM-Newton}}/OM 
observations. Masses are in units of solar mass,
$M_{\odot} = 1.99 \times 10^{33}$ g; 
temperature is in K; luminosity is in units of solar 
bolometric luminosity, $L_{\rm bol, \odot} = 3.9 \times 10^{33}$ 
erg s$^{-1}$. The observed colours constrain the temperature range 
(red lines). The observed brightness constrains the bolometric 
luminosity, as a function of temperature (blue lines).}  
\label{rsoria-C1_fig:fig5}
\end{figure}

It has been suggested (\cite{rsoria-C1:co83}; \cite{rsoria-C1:ne02}) 
that the companion is a main sequence B3 or B2.5 star 
($M_2$ \gtae~$7$ M$_{\odot}$). Such a star could 
also have the observed absolute magnitude 
$M_V = -1.21 \pm 0.16$ but would have  
much higher bolometric luminosity and temperature (bluer colors). 
However, their observations were carried out when the surface of the companion 
star was stronly heated by the X-ray source, and are therefore 
less reliable than our {\it XMM-Newton}/OM observation 
for a spectral classification.

The mean mass density in the Roche lobe of the companion star 
is uniquely determined by the binary period
(\cite{rsoria-C1:fr92}):
\begin{equation}
\rho \equiv \frac{3M_2}{4 \pi R_{\rm L}^3} \approx 115 P^{-2}_{\rm hr} 
\approx 0.069 \ \rm{g\  cm}^{-3}.
\end{equation}
We plot in Figure 6  
the evolutionary tracks in the ($M_{V}$, $\rho$) plane for 
a typical LMC metallicity $Z=0.008$ (eg, \cite{rsoria-C1:ca99}), 
compared with the mean density inside the Roche lobe. The dashed line 
corresponds to a radius of $0.95 R_{\rm L}$.
Stars with a mass $M \approx 4.5$ M$_{\odot}$ would be very close 
to filling their Roche lobe. 
Hence, mass transfer would occur mainly via Roche lobe overflow, 
in agreement with our X-ray observations. 
A more massive companion would not fill its Roche lobe. 
In particular, a B3V companion 
would only fill less than half of the volume of its Roche lobe. 
In that case, the mechanism 
of mass transfer would have to be a stellar wind. 
This is ruled out by the UV/optical colours and by the low column 
density inferred from the RGS data.

\begin{figure}
  \begin{center}
    \epsfig{file=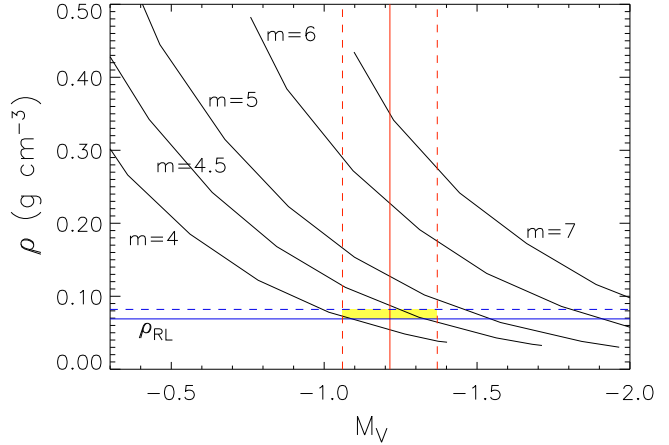, width=6.3cm, angle=270}
  \end{center}
\caption{We compare the mean density inside the RL 
of the secondary star ($\rho = 0.069$ g cm$^{-3}$, 
derived from the binary period) 
with the density of stars of various masses, 
in the ($M_V$,\,$\rho$) plane (evolutionary tracks 
from Girardi et al.\ 2000, assuming Z = 0.008). 
The dashed horizontal line corresponds to a radius 
$R = 0.95 R_{\rm L}$ for which Roche-lobe overflow 
becomes significant.
Only masses $M$ \ltae~$4.5$ M$_{\odot}$ are consistent 
with the observed brightness $M_V = -1.21 \pm 0.16$.}  
\label{rsoria-C1_fig:fig6}
\end{figure}

\section{Conclusions}

We have shown that the X-ray spectrum of LMC X-3 in the low-hard 
state is consistent with a simple power law. As the system returns to 
the high-soft state, the disk-blackbody component (interpreted 
as emission from an accretion disk) becomes more prominent. 
The luminosity in the low-hard state was 
$L_{\rm x} \approx 5 \times 10^{35}$~erg~s$^{-1}$ (0.3--10.0~keV band). 
In the high-soft state, 
$L_{\rm x} \approx 6 \times 10^{38}$~erg~s$^{-1}$, 
higher than the Eddington luminosity limit of an accreting 
neutron star, and consistent with the classification of LMC X-3 as a BHC.

The low line-of-sight column density, even at such a high luminosity, 
rules out wind accretion from a massive companion, and requires
that mass is transferred via RL overflow. In turns, this implies 
that the companion star is close to filling its RL.

We have shown that the optical/UV 
brightness and colours of the companion star suggest 
that it is a slightly evolved B5 star of mass $M_2 \approx 4.5$ 
M$_{\odot}$, rather than a main-sequence B3 star as previously thought. 
We have also shown that an evolved B5 companion would fill its RL, 
while a main-sequence B3 companion would not.

\begin{acknowledgements}

 We thank Keith Mason for his 
     comments.

\end{acknowledgements}


\begin{thebibliography}{}


%\bibitem[\protect\astroncite{Boyd \& Smale}{2000}]{rsoria-C1:bo00}
%      Boyd, P. T., Smale, A. P., 2000, IAUC 7424

\bibitem[\protect\astroncite{Boyd et~al.}{2000}]{rsoria-C1:bo00}
  Boyd, P. T., Smale, A. P., Homan, J., Jonker, P. G., 
    van der Klis, M., Kuulkers, E. 2000, ApJ, 542, L127

\bibitem[\protect\astroncite{Brandt \& Schulz}{2000}]{rsoria-C1:br00}
Brandt, W. N., Schulz, N. S. 2000, ApJ, 544, 123

\bibitem[\protect\astroncite{Caputo et~al.}{1999}]{rsoria-C1:ca99}
      Caputo, M., Marconi, G.,  Ripepi, V., 1999, ApJ 525, 784

\bibitem[\protect\astroncite{Cowley et~al.}{1983}]{rsoria-C1:co83}
      Cowley, A. P., Crampton, D., Hutchings, J. B., 
	Remillard, R., Penfold, J. E. 1983, ApJ 272, 118

\bibitem[\protect\astroncite{Frank et~al.}{1992}]{rsoria-C1:fr92}
      Frank, J., King, A., Raine, D. 1992, Accretion Power 
	in Astrophysics (Cambridge: University Press)

\bibitem[\protect\astroncite{Giles et~al.}{1986}]{rsoria-C1:gi86}
      Giles, D. R., Bolton, C. T., 1986, ApJ 304, 37

\bibitem[\protect\astroncite{Girardi et~al.}{2000}]{rsoria-C1:gi00}
      Girardi, L., Bressan, A., Bertelli, G., Chiosi, C., 2000, 
	A\&AS 141, 371

\bibitem[\protect\astroncite{Leong et~al.}{1971}]{rsoria-C1:le71}
  Leong, C., Kellogg, E., Gursky, H., Tanabaum, H., Giaccon, R. 1971, 
    ApJ, 170, L67

%\bibitem[\protect\astroncite{Mason et~al.}{1986}]{rsoria-C1:ma86}
%        Mason, K. O., Cropper, M. S., Hunt, R., Horner, S. D., 
%	Priedhorsky, W. C., Ho, C., Cordova, F. A., Jamar, C. A., 
%	Antonello, E., 1986, Proc. SPIE Vol.\ 2808 (eds: 
%	O. H. Siegmund, M. A. Gummin), 438

\bibitem[\protect\astroncite{Negueruela \& Coe}{2002}]{rsoria-C1:ne02}
        Negueruela, I., Coe, M. J., 2002, A\&A, in press (astro-ph/0201451)

\bibitem[\protect\astroncite{Paczy\'{n}ski}{1983}]{rsoria-C1:pa83}
       Paczy\'{n}ski, B. 1983, ApJ 273, L81

\bibitem[\protect\astroncite{Soria et~al.}{2001}]{rsoria-C1:so01}
       Soria, R., Wu, K., Page, M. J., Sakelliou, I., 
       2001, A\&A, 365, 273
 
\bibitem[\protect\astroncite{Sunyaev \& Titarchuk}{1980}]{rsoria-C1:su00}
       Sunyaev, R. A., Titarchuk, L. G. 1980, A\&A, 86, 121

\bibitem[\protect\astroncite{Titarchuk \& Zannias}{1998}]{rsoria-C1:ti98}
         Titarchuk, L. G., Zannias, T. 1998, ApJ, 193, 863

\bibitem[\protect\astroncite{van der Klis et~al.}{1985}]{rsoria-C1:va85}
        van der Klis, M., Clausen, J. V., Jensen, K., Tjemkes, S., 
	van Paradijs, J., 1985, A\&A 151, 322

\bibitem[\protect\astroncite{van Paradijs et~al.}{1987}]{rsoria-C1:va87}
        van Paradijs, J., van der Klis, M.,
 	Augusteijn, T., Charles, P., Corbet, R. H. D., 
	Ilovaisky, S., Maraschi, L., Motch, C., Pakull, M.,
 	Smale, A. P., Treves, A., van Amerongen, S., 1987, A\&A 184, 201


\bibitem[\protect\astroncite{Wilms et~al.}{2001}]{rsoria-C1:wi01}
        Wilms, J., Nowak, M. A., Pottschmidt, K., Heindl, W. A., 
	Dove, J. B., Begelman, M. C., 2001, MNRAS, 320, 327

\bibitem[\protect\astroncite{Wu et~al.}{2001}]{rsoria-C1:wu01}
        Wu, K., Soria, R., Page, M. J., Sakelliou, I., 
	Kahn, S. M., de Vries, C. P., 2001, 
	A\&A, 365, 267


\end{thebibliography}
\end{document}